\def\Snorm{\ton{\kap\bds\cdot\bds{\hat{h}}}}
\def\Sm{\ton{\kap\bds\cdot\bds{\hat{m}}}}
\def\Sl{\ton{\kap\bds\cdot\bds{\hat{l}}}}

\def\cu{\cos u}
\def\su{\sin u}

\def\rfr#1{equation (\ref{#1})}

\def\derp#1#2{\rp{\partial{#1}}{\partial{#2}}}
\def\dert#1#2{\frac{{{\textrm{d}}}{#1}}{{{\textrm{d}}}{#2}}}

\def\virg#1{`#1'}

\def\eqi{\begin{equation}}
\def\eqf{\end{equation}}
\def\eqia{\begin{eqnarray}}
\def\eqfa{\end{eqnarray}}

\def\rp#1#2{{#1\over#2}}
\def\lb#1{\label{#1}}
\def\kap{\bds{\hat{S}}}

\def\bds#1{\boldsymbol{#1}}

\def\co{\cos\omega}
\def\so{\sin\omega}

\def\cO{\cos\Omega}
\def\sO{\sin\Omega}

\def\cI{\cos I}
\def\sI{\sin I}

\def\ton#1{\left(#1\right)}
\def\qua#1{\left[#1\right]}
\def\grf#1{\left\{#1\right\}}

\documentclass[useAMS,usenatbib]{mn2e}
\usepackage{amsmath,textgreek,w-greek,wasysym}
\usepackage{amscd}
\usepackage{amssymb}
\usepackage{graphicx,epsfig}
\usepackage{txfonts}
\usepackage{xr-hyper}

\RequirePackage{color}

\allowdisplaybreaks[1]

\title[Post-Keplerian orbital periods]{Post-Keplerian corrections to the orbital periods of a two-body system and their measurability}

\author[L. Iorio]{L.
Iorio$^{1}$\thanks{E-mail:
lorenzo.iorio@libero.it}\\
$^{1}$I Ministero dell'Istruzione, dell'Universit\`{a} e della
Ricerca (M.I.U.R.), Viale Unit\`{a} di Italia 68
Bari, (BA) 70125,
Italy}

\begin{document}

\maketitle

\label{firstpage}

\begin{abstract}
The orbital motion of a binary system is characterized by various characteristic temporal intervals which, by definition, are different from each other: the draconitic, anomalistic and sidereal periods $T_\textrm{dra},~T_\textrm{ano},~T_\textrm{sid}$. They all coincide in the Keplerian case. Such a degeneracy is removed, in general, when a post-Keplerian acceleration is present. We analytically work out the corrections $T^{\ton{\textrm{pK}}}$ to such otherwise Keplerian periods which are induced by general relativity (Schwarzschild and Lense-Thirring) and, at the Newtonian level, by the quadrupole $J_2$ of the primary. In many astronomical and astrophysical systems, like exoplanets, one of the most accurately determined quantities is just the time span $T_\textrm{exp}$ characterizing the orbital revolution, which is often measured independently with different techniques like the transit photometry and the radial velocities.
Thus,  our results could be useful, in principle,  to either constrain the physical properties of the central body and/or perform new tests of general relativity, especially when no other standard observables like, e.g., the orbital precessions are accessible to observations. The difference $\Delta T$ of two independently measured periods would cancel out the common Keplerian term $T_\textrm{K}$ leaving just a post-Keplerian correction. Furthermore, by comparing the theoretically predicted post-Keplerian expressions $T^{\ton{\textrm{pK}}}$ with the experimental accuracy $\sigma_{T_\textrm{exp}}$ in measuring the orbital period(s) it is possible to identify those systems whose observations should be re-processed with genuine post-Keplerian models if $T^{\ton{\textrm{pK}}}>\sigma_{T_\textrm{exp}}$. It seems just the case for WASP-33 b since $\sigma_{T_\textrm{exp}}=0.04~\textrm{s}$, while $3~\textrm{s}\leq T_\textrm{dra}^{\ton{J_2}}\leq 9.5~\textrm{s},~T_\textrm{dra}^{\ton{\textrm{G\textcolor{black}{R}}}}=0.36~\textrm{s}$.
\end{abstract}


%

\begin{keywords}
general--celestial mechanics--ephemerides--gravitation
\end{keywords}
\section{Introduction}
From a theoretical point of view, there are many types of temporal intervals  characterizing different cyclic patterns in the orbital revolution of a binary system: the draconitic\footnote{This adjective was originally referred to the Moon's passage at its ascending node, when an eclipse occurs. Indeed, the ancient Greeks thought that, during an eclipse, our natural satellite was swallowed up by a dragon (\virg{\updelta\uprho$\acute{\upalpha}$\upkappa\upomega\upnu}, meaning literally \virg{which stares}) hiding near the nodes of the lunar orbit \citep{Capde05}. } period, which refers to two consecutive crossings of the ascending node; the anomalistic\footnote{The three anomalies are all zero (modulo 2\uppi) at the pericenter \citep{Capde05}. }  period, which characterizes the return at the periapsis; the sidereal period, i.e. the time required to describe a full revolution with respect to the fixed stars.

In a Keplerian two-body scenario, i.e. by neglecting any departures from spherical symmetry of the massive primary and the smaller revolving body under consideration and remaining within the Newtonian framework, all these three orbital periods coincide. It \textcolor{black}{does not happen} if the oblateness of the host star and\textcolor{black}{, e.g.,} general relativity  are taken into account along with their non-central additional accelerations with respect to the Newtonian monopole.

In several exoplanetary systems, one of the directly measured quantities determined with the highest accuracy is the time span $T_\textrm{exp}$ characterizing  the orbital revolution, generally dubbed as \virg{orbital period} \textcolor{black}{(see Section \ref{tran}) }. It is so because in most of them the planets orbit very closely to their parent stars; a huge number of full revolutions can be observed, with different techniques, over intervals even a few years long.
Thus, given the extraordinary accuracy with which the characteristic orbital time intervals are nowadays measured for the majority of the fast revolving exoplanets, a proper modeling of their post-Keplerian modifications may represent a further, independent valuable tool to either characterize the physical properties of the host stars and/or perform new tests of the gravitational theory by Einstein.
In the following, we will focus on transiting exoplanets, whose orbital periods are nowadays known at a  $10^{-7}-10^{-8}~\textrm{d}= 10^{-2}-10^{-3}~\textrm{s}$ accuracy level \citep{2013A&A...549A..10H, 2013ApJ...774...54S}. We will work in a first-order perturbation scheme to deal with the the non-central post-Keplerian accelerations of either classical and relativistic origin.

In Sections \ref{dracoper} to \ref{sidper}, we will set up a general method to consistently \textcolor{black}{calculate in Section \ref{PikeP} } the post-Keplerian corrections to  the draconitic, anomalistic and sidereal periods induced by any small perturbing acceleration with respect to the Newtonian monopole, irrespectively of its physical origin.
The potential application of such results to some specific astronomical scenarios is explored in Section \ref{uso}.
Section \ref{conclusioni} is devoted to the conclusions.
\subsection*{Notations}
Here,  basic notations and definitions used in the text are presented \citep{1991ercm.book.....B, 2000Monte}.
\begin{description}
\item[] $G:$ Newtonian constant of gravitation
\item[] $c:$ speed of light in vacuum
\item[] $M:$ mass of the primary
\item[] $\mu=GM:$ gravitational parameter of the primary
\item[] $S:$ angular momentum of the primary
\item[] ${\kap}:$ unit vector of the spin axis of the primary
\item[] $R:$ equatorial radius of the primary
\item[] $J_2:$ dimensionless quadrupole mass moment of the primary
\item[] $a:$  semimajor axis
\item[] $n_{\rm K} = \sqrt{\mu a^{-3}}:$   Keplerian mean motion
\item[] $T_{\rm K} = 2\uppi n_{\rm K}^{-1}:$ Keplerian orbital period
\item[] $e:$  eccentricity
\item[] $p=a(1-e^2):$  semilatus rectum
\item[] $I:$  inclination of the orbital plane
\item[] $\Omega:$  longitude of the ascending node
\item[] $\omega:$  argument of pericenter
\item[] $q=e\co:$  nonsingular orbital element $q$
\item[] $k=e\so:$  nonsingular orbital element $k$
\item[] $\bds{\hat{l}}=\grf{\cO,~\sO,~0}:$ unit vector directed along the line of the nodes toward the ascending node
\item[] $\bds{\hat{m}}=\grf{-\cI\sO,~\cI\cO,~\sI}:$ unit vector directed transversely to the line of the nodes in the orbital plane
\item[] $\bds{\hat{P}}=\bds{\hat{l}}\co +\bds{\hat{m}}\so:$ unit vector directed along the line of the apsides toward the pericenter
\item[] $\bds{\hat{Q}}= -\bds{\hat{l}}\so +\bds{\hat{m}}\co:$ unit vector directed transversely to the line of the apsides in the orbital plane
\item[] $f:$  true anomaly
\item[] $u=\omega + f:$  argument of latitude
\item[] ${\bds r} = r \ton{\bds{\hat{P}}\cos f +\bds{\hat{Q}}\sin f}:$ position vector of the test particle in terms of $f$
\item[] ${\bds r} = r \ton{\textcolor{black}{\bds{\hat{l}}}\cos u + \textcolor{black}{\bds{\hat{m}}}\sin u}:$ position vector of the test particle in terms of $u$
\item[] $r = p\ton{1 + e\cos f}^{-1}:$ distance of the test particle from the primary in terms of $f$
\item[] $r = p\ton{1 + q\cos u + k\sin u}^{-1}:$ distance of the test particle from the primary in terms of $u$
\item[] ${\bds v}=\sqrt{\mu p^{-1}}\qua{-\bds{\hat{P}}\sin f + \bds{\hat{Q}}\ton{\cos f + e}}:$ velocity vector of the test particle in terms of $f$
\item[] ${\bds v}=\sqrt{\mu p^{-1}}\qua{-\textcolor{black}{\bds{\hat{l}}}\ton{\sin u + k} + \textcolor{black}{\bds{\hat{m}}}\ton{\cos u + q} }:$ velocity vector of the test particle in terms of $u$
\item[] $\bds{\hat{r}}=\textcolor{black}{\bds{\hat{l}}}\cos u + \textcolor{black}{\bds{\hat{m}}}\sin u:$ radial unit vector
\item[] $\bds{\hat{h}}=\grf{\sI\sO,~-\sI\cO,~ \cI}:$ unit vector of the orbital angular momentum per unit mass of the test particle
\item[] $\bds{\hat{t}}= \bds{\hat{h}}\bds\times\bds{\hat{r}}= -\textcolor{black}{\bds{\hat{l}}}\sin u + \textcolor{black}{\bds{\hat{m}}}\cos u:$ transverse unit vector
\item[] $\varpi=\Omega+\omega:$ longitude of pericenter
\item[] $l=\varpi+f:$ true longitude
\item[] $Q=e\cos\varpi:$  nonsingular orbital element $Q$
\item[] $K=e\sin\varpi:$  nonsingular orbital element $K$
\item[] $r = p\ton{1 + Q\cos l + K\sin l}^{-1}:$ distance of the test particle from the primary in terms of $l$
\item[] $\mathcal{M}:$ mean anomaly
\item[] $\lambda=\varpi+\mathcal{M}:$ mean longitude
\item[] $\bds A:$ disturbing acceleration
\item[] $A_\textrm{R}=\bds A\bds\cdot\bds{\hat{r}}:$ radial component of $\bds A$
\item[] $A_\textrm{T}=\bds A\bds\cdot\bds{\hat{t}}:$ transverse component of $\bds A$
\item[] $A_\textrm{N}=\bds A\bds\cdot\bds{\hat{h}}:$ normal component of $\bds A$
\item[] $T_\textrm{dra}:$ draconitic period
\item[] $T_\textrm{ano}:$ anomalistic period
\item[] $T_{\rm sid}:$ sidereal period
\end{description}
\section{The orbital period(s) measured in extrasolar systems}\lb{tran}
In transiting exoplanets, the orbital period $T_\textrm{exp}$  which is actually measured is the time interval $T_{\rm tra}$ between two consecutive passages at the positions in the orbit, called transit centers, which minimize the sky-projected distance $r_{\rm sky}$ of the planet from the star \citep{Winn010}. By customarily assuming the plane of the sky as reference $\grf{x,~y}$ plane, in general, it is
\begin{align} r_{\rm sky} \lb{dxy} \nonumber &= \sqrt{x^2 + y^2} = \rp{p\sqrt{3+\cos \ton{2\omega + 2f} +2\cos 2 I\sin^2\ton{\omega+f}}}{2\ton{1+e\cos f}}=\\ \nonumber \\
& = \rp{a}{2}\sqrt{3+\cos 2u +2\cos 2 I\sin^2 u}+\mathcal{O}\ton{e}.
\end{align}
In obtaining \rfr{dxy}, we did not make any a-priori assumption about the orientation of the orbital plane in space; cfr. with Eq. (5) of \citet{Winn010} obtained by setting $\Omega=180^{\circ}$.
From \rfr{dxy}, it turns out that, to zero order in eccentricity,  $T_{\rm tra}$ is the time interval from $u=u^{\ast}$ to $u=u^{\ast}+2\uppi$, where $u^{\ast}$ is the specific value corresponding to the transit; it yields a local minimum of $r_{\rm sky}$ \citep{Winn010}, but its actual value is not of strict importance here. Thus, for circular orbits, it seems  reasonable to infer \eqi T_{\rm tra}= T_{\rm dra},\eqf i.e. it should be possible to identify the orbital period actually measured for transiting exoplanets with their draconitic period because of the argument of latitude $u$ entering \rfr{dxy}, which is reckoned just from the line of the nodes. In the general case of an elliptical orbit, also the anomalistic period may come into play in view of $f$ in the denominator of \rfr{dxy}.

The radial velocity $v_r$ is another observable widely adopted in the field of extrasolar planetary systems; it is used also for some transiting planets themselves. Since it is \citep{Bat01}
\eqi v_r \textcolor{black}{\propto} e\cos\omega + \cos\ton{\omega + f},\eqf similar considerations as before should hold: while in the circular orbit approximation the orbital period measured from the radial velocity curve should be considered as a draconitic one, in the general case it may be identified with the anomalistic period $T_\textrm{ano}$.
\section{The draconitic period in a post-Keplerian orbit}\lb{dracoper}
The draconitic period $T_\textrm{dra}$, defined for a perturbed trajectory as  the time interval between two successive instants when the real position of the test particle coincides with the ascending node position on the corresponding osculating orbit, can be calculated as \citep{1977AN....298..107M}
\eqi T_\textrm{dra} = \int_0^{2\uppi}\ton{\dert t u} \textrm{d}u.\lb{Tdracon}\eqf
In a general post-Keplerian scenario, from the definition of the argument of latitude $u$, it follows
\eqi\dert u t =\dert\omega t + \dert f t.\lb{pimpa}\eqf
In it, \citep{1958SvA.....2..147E, 1991ercm.book.....B, 2003ASSL..293.....B,2014PhRvD..89d4043W}
\eqi\dert f t  = \rp{\sqrt{\mu p}}{r^2} - \dert\omega t -\cI\dert\Omega t; \lb{anous}\eqf full details can be found in \citet{1958SvA.....2..147E}.  Thus,
\textcolor{black}{eq. \ref{pimpa}} can be written
\eqi \dert u t = \rp{\sqrt{\mu p}}{r^2}\qua{1 - \rp{r^2\cI}{\sqrt{\mu p}}\dert\Omega t},\eqf
 so that \citep{Ocho59, 1977AN....298..107M}
\eqi\dert t u = \rp{r^2\upalpha}{\sqrt{\mu p}},\lb{dtdu}\eqf
in which the definition
\eqi\upalpha\doteq \rp{1}{1 - \rp{r^2\cI}{\sqrt{\mu p}}\dert\Omega t }\eqf
is adopted.
To the first order in the disturbing acceleration entering $\textrm{d}\Omega/\textrm{d}t$ through \citep{Ocho59, 1977AN....298..107M}
\eqi\dert\Omega u = \rp{r^3\upalpha\sin u A_\textrm{N}}{\mu p\sin I},\lb{dOdu}\eqf
\rfr{dtdu} can be expanded as
\eqi\dert t u \simeq \rp{r^2}{\sqrt{\mu p}} + \rp{r^4\cI}{\mu p}\dert\Omega t.\lb{dtdu2}\eqf

When a disturbing acceleration $\bds A$ is present, it affects $dt/du$ in a twofold way.
A direct change with respect to the Keplerian case arises from the node rate, while further contributions come  also from the first term in \rfr{dtdu2} when the instantaneous shifts of the orbital elements $\grf{\xi}$ entering it are properly taken into account; in the notation of \citet{1981CeMec..23...83R}, they can be dubbed as \virg{indirect}.  More specifically, from the expression of $r\ton{u}$,
by posing
\eqi F\ton{p,~q,~k}\doteq \rp{r^2}{\sqrt{\mu p}},\eqf it is
\eqi F= \left. F\right|_{\rm K} +\Delta F=\left. F\right|_{\rm K} + \sum_\xi^{p,~q,~k}\left.\derp F \xi\right|_{\rm K}\Delta\xi\ton{u_0,~u},\lb{expa}\eqf
where the subscript $\virg{\textrm{K}}$ refers to the unperturbed, Keplerian ellipse, and $\xi = p,~q,~k$.

As a consequence of both the indirect and direct contributions, the draconitic time lapse can be analytically computed from \rfr{Tdracon}, with \rfr{dtdu2} and \rfr{expa}, as \citep{1977AN....298..107M}
\eqi T^{\ton{\textrm{pK}}}_\textrm{dra} = T_\textrm{dra} - T_{\rm K} = I_1^\textrm{dra}+I_2^\textrm{dra}+I_3^\textrm{dra}+I_4^\textrm{dra}, \eqf
with
\begin{align}
I_1^\textrm{dra} \lb{I1dr} \nonumber & = \int_0^{2\uppi}\left.\derp{F}{p}\right|_{\rm K}\Delta p\ton{u_0,~u} \textrm{d}u=\\ \nonumber \\
& = \rp{3}{2}\sqrt{\rp{p}{\mu}}\int_0^{2\uppi}\rp{\Delta p\ton{u_0,~u} }{\ton{1+q\cu+k\su}^2}\textrm{d}u, \\ \nonumber \\
I_2^\textrm{dra} \nonumber & = \int_0^{2\uppi}\left.\derp{F}{q}\right|_{\rm K}\Delta q\ton{u_0,~u} \textrm{d}u=\\ \nonumber \\
& = -2\sqrt{\rp{p^3}{\mu}}\int_0^{2\uppi}\rp{\Delta q\ton{u_0,~u}\cu }{\ton{1+q\cu+k\su}^3}\textrm{d}u, \\ \nonumber \\
I_3^\textrm{dra} \lb{I3dr} \nonumber & = \int_0^{2\uppi}\left.\derp{F}{k}\right|_{\rm K}\Delta k\ton{u_0,~u} \textrm{d}u = \\ \nonumber \\
& = -2\sqrt{\rp{p^3}{\mu}}\int_0^{2\uppi}\rp{\Delta k\ton{u_0,~u}\su }{\ton{1+q\cu+k\su}^3}\textrm{d}u, \\ \nonumber \\
I_4^\textrm{dra} \lb{I4dr} & = \int_0^{2\uppi}\rp{r^4 \cos I}{\mu p}\dert\Omega t \textrm{d}u.
\end{align}
In equations (\ref{I1dr})--(\ref{I3dr}), the instantaneous shifts of $p,~q,~k$
are to be calculated as \citep{1977AN....298..107M}
\begin{align}
\Delta p\ton{u_0,~u} &=\int_{u_0}^u\ton{\dert p  {u^{'}}}\textrm{d}u^{'}, \\ \nonumber \\
\Delta q\ton{u_0,~u} &=\int_{u_0}^u\ton{\dert q  {u^{'}}}\textrm{d}u^{'}, \\ \nonumber \\
\Delta k\ton{u_0,~u} &=\int_{u_0}^u\ton{\dert k  {u^{'}}}\textrm{d}u^{'},
\end{align}
by using the analytical expressions \citep{Ocho59, 1977AN....298..107M}
\begin{align}
\dert p u & = \rp{2r^3\upalpha A_\textrm{T}}{\mu}, \\ \nonumber \\
\dert q u \nonumber & = \rp{r^3 \upalpha~k \su \cot I A_\textrm{N}}{\mu p} + \\ \nonumber \\
\nonumber & + \rp{r^2\upalpha\qua{\rp{r}{p}\ton{q + \cu}  + \cu}A_\textrm{T}}{\mu} + \\ \nonumber \\
& + \rp{r^2\upalpha\su A_\textrm{R}}{\mu}, \\ \nonumber\\
\dert k u \nonumber & = -\rp{r^3 \upalpha~q \su \cot I A_\textrm{N}}{\mu p} + \\ \nonumber \\
\nonumber & + \rp{r^2\upalpha\qua{\rp{r}{p}\ton{k + \su}  + \su}A_\textrm{T}}{\mu} - \\ \nonumber \\
& - \rp{r^2\upalpha\cu~A_\textrm{R}}{\mu}.
\end{align}
of the derivatives of $p,~q,~k$ with respect to $u$ to the first order in the disturbing acceleration, i.e. by setting $\upalpha = 1$.
The time derivative of the node entering \rfr{I4dr} has to be calculated, to the first order in the perturbation, from \rfr{dOdu} and \rfr{dtdu}.
It is intended that the right-hand-sides of equations (\ref{I1dr})--(\ref{I4dr}) are evaluated onto the unperturbed Keplerian ellipse.
\section{The anomalistic period in a post-Keplerian orbit}\lb{anoper}
The anomalistic period $T_\textrm{ano}$, defined as  the time interval between two successive instants when the real position of the test particle coincides with the pericenter position on the corresponding  orbit, can be calculated as \citep{Zhongo60, 1979AN....300..313M}
\eqi T_\textrm{ano} = \int_0^{2\uppi}\ton{\dert t f} \textrm{d}f.\eqf
In presence of a disturbing acceleration $\bds A$, one has \citep{1959ForPh...7S..55T, 1979AN....300..313M}
\begin{align}
\dert t f \lb{dtdf} &= \rp{r^2\upbeta}{\sqrt{\mu p}},\\ \nonumber \\
\upbeta & \doteq \rp{1}{1-  \rp{r^2}{\sqrt{\mu p}}\ton{\dert\omega t+\cI\dert\Omega t} }. \lb{beta}
\end{align}
The anomalistic period $T_\textrm{ano}$ generally differs from $T_\textrm{K}$  by an additive post-Keplerian term $T_\textrm{ano}^{\ton{\textrm{pK}}}$ which can be analytically worked out as \citep{1979AN....300..313M}
\eqi T^{\ton{\textrm{pK}}}_\textrm{ano} \lb{TanoLT} = T_\textrm{ano} - T_\textrm{K} = I_1^\textrm{ano}+I_2^\textrm{ano}+I_3^\textrm{ano}, \eqf
with \citep{Zhongo60, 1979AN....300..313M}
\begin{align}
I_1^\textrm{ano} \lb{I1an} & =\rp{3}{2}\sqrt{\rp{p}{\mu}}\int_0^{2\uppi} \rp{\Delta p\ton{f_0,~f}}{\ton{1+e\cos f}^2}\textrm{d}f, \\ \nonumber \\
I_2^\textrm{ano} \lb{I2an} & = -2\sqrt{\rp{p^3}{\mu}}\int_0^{2\uppi} \rp{\Delta e\ton{f_0,~f}\cos f}{\ton{1+e\cos f}^3}\textrm{d}f, \\ \nonumber \\
I_3^\textrm{ano} \lb{I3an} & =\int_0^{2\uppi}\rp{r^4}{\mu p}\ton{\dert\omega t + \cI\dert\Omega t}\textrm{d}f.
\end{align}
In equations (\ref{I1an})--(\ref{I2an}), the instantaneous shifts of $p,e$
are to be calculated as \citep{1979AN....300..313M}
\begin{align}
\Delta p\ton{f_0,~f} &=\int_{f_0}^f\ton{\dert p  {f^{'}}}\textrm{d}f^{'}, \\ \nonumber \\
\Delta e\ton{f_0,~f} &=\int_{f_0}^f\ton{\dert e  {f^{'}}}\textrm{d}f^{'},
\end{align}
by using the analytical expressions \citep{1959ForPh...7S..55T, 1979AN....300..313M}
\begin{align}
\dert p f & = \rp{2r^3\upbeta A_\textrm{T}}{\mu}, \\ \nonumber \\
\dert e f & = \rp{r^2\upbeta}{\mu}\qua{\sin f A_\textrm{R} + \ton{1+\rp{r}{p}}\cos f A_\textrm{T}  +e\ton{\rp{r}{p}}A_\textrm{T}}
\end{align}
of the derivatives of $p,e$ with respect to $f$ to the first order in the disturbing acceleration, i.e. by setting $\upbeta = 1$.
Moreover, it is intended that the right-hand-sides of equations (\ref{I1an})--(\ref{I3an}) are evaluated onto the unperturbed Keplerian ellipse. In particular, the time derivatives entering \rfr{I3an} has to be calculated, to the first order in the perturbation, from \citep{1959ForPh...7S..55T, 1979AN....300..313M}
\begin{align}
\dert\Omega f &= \rp{r^3\upbeta \sin\ton{\omega + f}A_\textrm{N}}{\mu p \sI}, \\ \nonumber \\
\dert\omega f \nonumber & = \rp{r^2\upbeta}{\mu}\qua{-\rp{\cos f~A_\textrm{R}}{e} +\ton{1+\rp{r}{p} }\rp{\sin f~A_\textrm{T}}{e} -\right.\\ \nonumber \\
&-\left. \ton{\rp{r}{p}}\cot I\sin\ton{\omega + f}A_\textrm{N} }
\end{align}
and \rfr{dtdf} with $\upbeta=1$.

An equivalent computational approach ca be found in \citet{1981CeMec..23...83R}; the expression for $\textrm{d}t_a^{(1)}/\textrm{d}\theta$ in Eq. (3) of \citet{1981CeMec..23...83R} is incorrect since an overall multiplicative factor $\sin\theta$ is missing in the second term of the right-hand-side. In the notation of \citet{1981CeMec..23...83R}, $\theta$ is the true anomaly.
\section{The sidereal period in a post-Keplerian orbit}\lb{sidper}
The sidereal period can be defined as the time interval between
two successive instants when the real position of the test particle
lies on a given reference direction in the sky. By assuming that the latter is the one from which the longitudes are reckoned, i.e. the $x$ axis in the coordinate systems which are usually tied to the plane of the sky, a plausible expression for the sidereal period can be calculated by means of the true longitude\footnote{In principle, also the mean longitude $\lambda$ might be used. Nonetheless, it may be less easily identified with actually measurable repeating temporal intervals. Furthermore, also the calculation would be more cumbersome because one should adopt the mean anomaly $\mathcal{M}$ as fast variable of integration, while most of the available analytical expressions for the instantaneous rates of the elements, etc. use the true anomaly $f$. } $l$ as
\eqi T_{\rm sid} \lb{Tsid} = \int_0^{2\uppi}\ton{\dert t l} \textrm{d}l,\eqf
in close analogy with  Sections \ref{dracoper} to \ref{anoper}.
From the definition of $l$, it follows
\eqi\dert l t =\dert\Omega t + \dert\omega t + \dert f t.\lb{gosh}\eqf
By using \rfr{anous}, \rfr{gosh} can be written
\eqi \dert l t = \rp{\sqrt{\mu p}}{r^2}\qua{1 + \rp{r^2\ton{1 - \cI}}{\sqrt{\mu p}}\dert\Omega t}.\eqf
Thus,
\eqi\dert t l = \rp{r^2\upgamma}{\sqrt{\mu p}},\lb{dtdl}\eqf
in which we define
\eqi\upgamma\doteq \rp{1}{1 + \rp{r^2\ton{1 - \cI}}{\sqrt{\mu p}}\dert\Omega t }.\eqf
To the first order in the disturbing acceleration entering $d\Omega/dt$, \rfr{dtdl} can be expanded as
\eqi\dert t l \simeq \rp{r^2}{\sqrt{\mu p}} + \rp{r^4\ton{\cI - 1}}{\mu p}\dert\Omega t.\lb{dtdl2}\eqf

In calculating \rfr{Tsid} through \rfr{dtdl2}, the second term of \rfr{dtdl2} with
\eqi
\dert\Omega t = \rp{r\upalpha\sin\ton{l-\Omega} A_\textrm{N}}{\sqrt{\mu p}\sI}\lb{dNdt}
\eqf
evaluated onto the unperturbed Keplerian ellipse, i.e. for $\upgamma=1$, yields the direct correction to the sidereal period.

As seen in Sections \ref{dracoper} to \ref{anoper}, further, indirect contributions arise  also from the first term in \rfr{dtdu2} when the instantaneous shifts of the orbital elements $\grf{\psi}$ entering it  are properly taken into account.  From $r\ton{l}$,
by posing
\eqi W\ton{p,~Q,~K}\doteq \rp{r^2}{\sqrt{\mu p}},\eqf it is
\eqi W= \left. W\right|_\textrm{K} +\Delta W = \left. W\right|_{\rm K}+\sum_\psi^{p,~Q,~K}\left.\derp W \psi\right|_{\rm K}\Delta\psi\ton{l_0,~l},\eqf
where  $\psi = p,~Q,~K$.

Thus, by accounting for both the indirect and the direct contributions, we have for the post-Keplerian correction to the sidereal period
\eqi T_{\rm sid}^{\ton{\textrm{pK}}} = T_{\rm sid} - T_{\rm K} =I_1^{\rm sid} + I_2^{\rm sid} + I_3^{\rm sid} + I_4^{\rm sid},\eqf
with
\begin{align}
I_1^{\rm sid} \lb{I1sid} \nonumber & = \int_0^{2\uppi}\left.\derp{W}{p}\right|_\textrm{K}\Delta p\ton{l_0,~l} \textrm{d}l\\ \nonumber \\
&= \rp{3}{2}\sqrt{\rp{p}{\mu}}\int_0^{2\uppi}\rp{\Delta p\ton{l_0,~l} }{\ton{1 + Q\cos l + K\sin l}^2}\textrm{d}l, \\ \nonumber \\
I_2^{\rm sid} \nonumber & = \int_0^{2\uppi}\left.\derp{W}{Q}\right|_\textrm{K}\Delta Q\ton{l_0,~l} \textrm{d}l\\ \nonumber \\
&=  -2\sqrt{\rp{p^3}{\mu}}\int_0^{2\uppi}\rp{\Delta Q\ton{l_0,~l}\cos l }{\ton{1 + Q\cos l + K\sin l}^3}\textrm{d}l, \\ \nonumber \\
I_3^{\rm sid} \lb{I3sid} \nonumber & = \int_0^{2\uppi}\left.\derp{W}{K}\right|_\textrm{K}\Delta K\ton{l_0,~l} \textrm{d}l\\ \nonumber \\
&=  -2\sqrt{\rp{p^3}{\mu}}\int_0^{2\uppi}\rp{\Delta K\ton{l_0,~l}\sin l }{\ton{1 + Q\cos l + K\sin l}^3}\textrm{d}l, \\ \nonumber \\
I_4^{\rm sid} \lb{I4sid} & = \int_0^{2\uppi}\rp{r^4 \ton{\cI-1}}{\mu p}\dert\Omega t \textrm{d}l.
\end{align}
In equations (\ref{I1sid})--(\ref{I4sid}), the instantaneous shifts of $p,~Q,~K$
are to be calculated as
\begin{align}
\Delta p\ton{l_0,~l} \lb{Dpl} &=\int_{l_0}^l\ton{\dert p  {l^{'}}}\textrm{d}l^{'}, \\ \nonumber \\
\Delta Q\ton{l_0,~l} &=\int_{l_0}^l\ton{\dert Q  {l^{'}}}\textrm{d}l^{'}, \\ \nonumber \\
\Delta K\ton{l_0,~l} \lb{DKl} &=\int_{l_0}^l\ton{\dert K  {l^{'}}}\textrm{d}l^{'},
\end{align}
by using the analytical expressions
\begin{align}
\dert p l \lb{dpdl}& = \rp{2r^3\upgamma A_\textrm{T}}{\mu}, \\ \nonumber \\
\dert Q l \nonumber & = -\rp{r^3 \upgamma~K \sin\ton{l-\Omega} \tan\ton{I/2}A_\textrm{N}}{\mu p} + \\ \nonumber \\
\nonumber & + \rp{r^2\upgamma\qua{r~Q + \ton{r + p}\cos l}A_\textrm{T}}{\mu p} +\\ \nonumber \\
& + \rp{r^2\upgamma\sin l~A_\textrm{R}}{\mu}, \\ \nonumber\\
\dert K l \lb{dKdl} \nonumber & = \rp{r^3 \upgamma~Q \sin\ton{l-\Omega} \tan\ton{I/2}A_\textrm{N}}{\mu p} + \\ \nonumber \\
\nonumber & + \rp{r^2\upgamma\qua{r~K + \ton{r + p}\sin l}A_\textrm{T}}{\mu p} -\\ \nonumber \\
& - \rp{r^2\upgamma\cos l~A_\textrm{R}}{\mu}.
\end{align}
to the first order in the disturbing acceleration, i.e. by setting $\upgamma = 1$.
It is intended that equations (\ref{I1sid})--(\ref{dKdl}) are evaluated onto the unperturbed Keplerian ellipse.
\section{The post-Keplerian corrections to the orbital periods}\label{PikeP}
In this Section, we look at  the Newtonian and post-Newtonian non-central accelerations resolving the degeneracy of the orbital periods with respect to the Keplerian case. In Section \ref{J2PER}, the impact of the quadrupole mass moment of the primary is considered to the Newtonian level. The general relativistic post-Newtonian accelerations are treated in Section \ref{SCHPER} (Schwarzschild) and Section \ref{LTPER} (Lense-Thirring).
\subsection{The Newtonian quadrupole correction to the orbital periods}\lb{J2PER}
The classical post-Keplerian acceleration felt by a test particle in the field of an oblate primary is \citep{2005CeMDA..91..217V}
\eqi{\bds A}^{\ton{J_2}}=\rp{3 J_2 \mu R^2}{2r^4}\grf{\qua{5\ton{\bds{\hat{S}}\bds\cdot\bds{\hat{r}}}^2-1}\bds{\hat{r}} - 2\ton{\bds{\hat{S}}\bds\cdot\bds{\hat{r}}}\bds{\hat{S}}}.\eqf
Its radial, transverse and normal components are
\begin{align}
A_\textrm{R}^{\ton{J_2}} & = \rp{3 J_2\mu R^2}{2 r^4}\qua{3\ton{\bds{\hat{S}}\bds\cdot\bds{\hat{r}}}^2 - 1}, \\ \nonumber \\
A_\textrm{T}^{\ton{J_2}}  &= -\rp{3 J_2\mu R^2}{r^4}\ton{\kap\bds\cdot\bds{\hat{r}}}\ton{\kap\bds\cdot\bds{\hat{t}}}, \\ \nonumber\\
A_\textrm{N}^{\ton{J_2}} & = -\rp{3 J_2\mu R^2}{r^4}\ton{\kap\bds\cdot\bds{\hat{r}}}\ton{\kap \bds\cdot\bds{\hat{h}}}.
\end{align}

To zero order in eccentricity, the draconitic, sidereal and anomalistic corrections due to $J_2$ are
\begin{align}
T_{\rm dra}^{\ton{J_2}}\lb{TdraJ2} &= \rp{3\uppi J_2 R^2}{2\sqrt{\mu a}}~{\mathcal{T}}_{\textrm{dra}}^{\ton{J_2}}\ton{I,~\Omega,~\kap}+\mathcal{O}\ton{e^n},~n\geq 1,\\ \nonumber \\
T_{\rm ano}^{\ton{J_2}} &= \rp{3\uppi J_2 R^2}{2\sqrt{\mu a}}~{\mathcal{T}}_{\textrm{ano}}^{\ton{J_2}}\ton{I,~\Omega,~\kap}+\mathcal{O}\ton{e^n},~n\geq 1,\\ \nonumber \\
T_{\rm sid}^{\ton{J_2}} &= \rp{3\uppi J_2 R^2}{2\sqrt{\mu a}}~{\mathcal{T}}_{\textrm{sid}}^{\ton{J_2}}\ton{I,~\Omega,~\kap}+\mathcal{O}\ton{e^n},~n\geq 1,\lb{TsidJ2}
\end{align}
with
\begin{align}
\mathcal{T}_{\textrm{dra}}^{\ton{J_2}} \lb{tdraJ2}\nonumber &= - 4 + 6\Sl\textcolor{black}{^2} + 6\Sm\textcolor{black}{^2} + \\ \nonumber \\
\nonumber & + 3\qua{\Sl^2-\Sm^2}\cos 2 u_0 + \\ \nonumber \\
\nonumber & +  6\Sl\Sm\sin 2u_0 - \\ \nonumber \\
& - 2\Snorm\Sm\cot I, \\ \nonumber\\
\mathcal{T}_{\textrm{ano}}^{\ton{J_2}} \lb{tanoJ2}\nonumber &= - 2 + 3\Sl\textcolor{black}{^2} + 3\Sm\textcolor{black}{^2} + \\ \nonumber \\
\nonumber & + 3\qua{\Sl^2-\Sm^2}\cos 2 u_0 + \\ \nonumber \\
          & +  6\Sl\Sm\sin 2u_0, \\ \nonumber \\
\mathcal{T}_{\textrm{sid}}^{\ton{J_2}} \lb{tsidJ2}\nonumber &= - 4 + 6\Sl\textcolor{black}{^2} + 6\Sm\textcolor{black}{^2} + \\ \nonumber \\
\nonumber & + 3\qua{\Sl^2-\Sm^2}\cos 2 u_0 + \\ \nonumber \\
\nonumber & +  6\Sl\Sm\sin 2u_0 + \\ \nonumber \\
& + 2\Snorm\Sm\tan\ton{\rp{I}{2}}.
\end{align}

Note that equations (\ref{TdraJ2})--(\ref{TsidJ2}) fall off as $1/\sqrt{a}$.
Contrary to the relativistic  corrections (see Sections \ref{SCHPER} to \ref{LTPER}), equations (\ref{tdraJ2})--(\ref{tsidJ2}) do depend, in general, on the initial position of the test particle along its orbit through $u_0$ already to zero order in eccentricity. Such a unique feature may be exploited either to enhance or reduce the magnitude of the post-Keplerian  component of the period(s) measured, at least to a certain extent, if $J_2$ is the target of the observational campaign or, vice versa, if it is viewed as a competing source of systematic uncertainty when other dynamical effects like GR are looked for.
\subsection{The 1pN corrections to the orbital periods for a non-rotating primary}\lb{SCHPER}
To the first post-Newtonian order (1pN), the Schwarzschild-type gravitoelectric \textcolor{black}{(GE)} acceleration due to a static mass is \citep{Sof89}
\eqi
{\bds A}^{(\rm GE)} = \rp{\mu}{c^2 r^2}\qua{\ton{\rp{4\mu}{r}-v^2 }\bds{\hat{r}}  + 4 \ton{\bds{\hat{r}}\bds\cdot\bds v}\bds v}.\lb{AGE}\eqf
%
%
%
%

The resulting corrections to the orbital periods are
\begin{align}
T_{\rm dra}^{\ton{\textrm{GE}}} &= \rp{12\uppi\sqrt{\mu a}}{c^2 }+\mathcal{O}\ton{e^n},~n\geq 1,\lb{TdraGE}\\ \nonumber \\
T_{\rm ano}^{\ton{\textrm{GE}}} &= \rp{3\uppi\sqrt{\mu a}}{c^2 \ton{1-e^2}^2}\mathcal{T}_\textrm{ano}^{\ton{\textrm{GE}}},\lb{TanoGE}\\ \nonumber \\
T_{\rm sid}^{\ton{\textrm{GE}}} &= \rp{12\uppi\sqrt{\mu a}}{c^2}+\mathcal{O}\ton{e^n},~n\geq 1,\lb{TsidGE}
\end{align}
with
\begin{align}
\mathcal{T}_\textrm{ano}^{\ton{\textrm{GE}}} &= 6 + 7 e^2 + 2 e^4 + 2 e \ton{7 + 3 e^2} \cos f_0 + 5 e^2 \cos 2 f_0.
\end{align}
They depend on the orbital size as $\sqrt{a}$, and depend on the initial orbital position along the orbit to $\mathcal{O}\ton{e^n},~n\geq 1$.

For other calculations of some of the periods treated here,  performed with different computational techniques and approximation schemes, see, e.g., \citet{1985AIHS...43..107D, 1986ApJ...305..623B, 1987CeMec..40...77S, Sof89, 1991ercm.book.....B, 2000ApJS..126...79G,2001PhLA..292...49M}.
\subsection{The 1pN gravitomagnetic  corrections to the orbital periods}\lb{LTPER}
The 1pN Lense-Thirring \textcolor{black}{(LT)} acceleration  experienced by a test particle orbiting  \textcolor{black}{a stationary source such as} a slowly rotating \textcolor{black}{body}  is \citep{Sof89}
\eqi
{\bds A}^{(\rm LT)}   = \rp{2GS}{c^2 r^3}\qua{3\ton{\kap\bds\cdot\bds{\hat{r}}}\bds{\hat{r}}\bds\times\bds v+ \bds v\bds\times\kap}\lb{LTacc}.
\eqf
For any values of the orbital and physical parameters characterizing known astronomical and astrophysical systems, \rfr{LTacc} can be considered as a small perturbation of the Newtonian monopole.

%
%
%
%
%

%
The gravitomagnetic contributions to the orbital periods considered, to zero order in eccentricity, turn out to be
\begin{align}
T_{\rm dra}^{\ton{\textrm{LT}}}\lb{TdraLT} &= \rp{4\uppi S}{c^2 M}{\mathcal{T}}_{\textrm{dra}}^{\ton{\textrm{LT}}}\ton{I,~\Omega,~\kap}+\mathcal{O}\ton{e^n},~n\geq 1,\\ \nonumber \\
T_{\rm ano}^{\ton{\textrm{LT}}} &= 0,\\ \nonumber \\
T_{\rm sid}^{\ton{\textrm{LT}}} &= \rp{4\uppi S}{c^2 M}{\mathcal{T}}_{\textrm{sid}}^{\ton{\textrm{LT}}}\ton{I,~\Omega,~\kap}+\mathcal{O}\ton{e^n},~n\geq 1,\lb{TsidLT}
\end{align}
with
\begin{align}
\mathcal{T}_{\textrm{dra}}^{\ton{\textrm{LT}}} \lb{tdraLT}&= 2\Snorm + \Sm\cot I, \\ \nonumber\\
\mathcal{T}_{\textrm{sid}}^{\ton{\textrm{LT}}} \lb{tsidLT}&=2\Snorm - \Sm\tan\ton{\rp{I}{2}}.
\end{align}
%
It is worthwhile noticing that equations (\ref{TdraLT})--(\ref{TsidLT}) are independent of both $G$ and  $a$. Moreover,  equations (\ref{tdraLT})--(\ref{tsidLT})   do not depend on the initial position of the test particle along its orbit.

For different calculations  of other gravitomagnetic characteristic orbital time spans present in the literature, see, e.g., \citet{2001PhLA..292...49M,2014PhRvD..90d4059H}.
\section{How to use the modeled post-Keplerian corrections to the orbital periods?}\lb{uso}
In a series of papers \citep{Zhongo66, Amelin66, Kassi66}, it was demonstrated that it is possible to measure the draconitic period of an artifical Earth's satellite\footnote{It was the Soviet spacecraft $1960-\varepsilon$ 3 for which simultaneous visual tracking data were used \citep{Zhongo66, Amelin66, Kassi66}.} as the ratio of the difference of the times of passages of the sub-satellite point through a chosen parallel for two following epochs to the number of satellite revolutions corresponding to this difference. The accuracy reached at that time seems to be of the order of $10^{-4}$ s \citep{Kassi66}; it is highly plausible that it could be improved by many orders of magnitude with the most recent techniques currently available.

In several practical Solar System scenarios, the tracking of spacecraft reaches its highest level of accuracy usually at their closest approaches to
their target bodies thanks to a variety of techniques such as Doppler ranging rate, laser ranging, etc. \citep{2007IJMPD..16.2117I,2009AcAau..65..666I}. Thus, it appears meaningful, at least in principle,  to look also at the anomalistic period.

As far as exoplanetary systems are concerned,
we can use our results to predict the expected magnitude of the post-Keplerian components $T^{\ton{\textrm{pK}}}$ of the orbital periods and compare them to the currently available experimental accuracy in measuring $T_\textrm{exp}$, to be identified with some of the orbital periods considered here. If $T^{\ton{\textrm{pK}}}>\sigma_{T_\textrm{exp}}$, a truly post-Keplerian fit of the observations is, actually, required; otherwise, the values of the system's parameters estimated with a purely Keplerian model  based on the 3rd Kepler's law should be considered as wrong since they would be biased by the non-negligible post-Keplerian effect(s). By looking at WASP-33 b \citep{2010MNRAS.407..507C}, Eq. \ref{TdraJ2} and Eq. \ref{tdraJ2}, computed with the currently available\footnote{In fact, they were derived in a Keplerian framework.} orbital and physical parameters \citep{2010MNRAS.407..507C, 2011MNRAS.416.2096S, 2015A&A...578L...4L} along with the orbital precession-induced value of $J_2$ \citep{2015ApJ...810L..23J, 2016MNRAS.455..207I}, yields an expected draconitic correction due to the stellar quadrupole ranging from 3 s to $9.5$ s, depending on $u_0$ assumed as a free parameter; it is almost two orders of magntitude larger than $\sigma_{T_\textrm{exp}}=4.5\times 10^{-7}~\textrm{d}=0.04~\textrm{s}$ \citep{2011MNRAS.416.2096S}. Moreover, from Eq. \ref{TdraGE}, it turns out that the Schwarzschild-like  draconitic period amounts to $0.36$ s, which is almost one order of magnitude larger than $\sigma_{T_\textrm{exp}}$. Thus, it can be concluded that the data record of WASP-33 b should be re-analyzed within a truly post-Keplerian scenario including both the stellar quadrupole at the Newtonian level and general relativity.

On the other hand, if, for a given system, it were  possible to have two independently measured orbital periods identifiable with some of the ones examined here, their difference $\Delta T$ would cancel out the Keplerian period $T_\textrm{K}$, leaving just a post-Keplerian correction due to the fact that the perturbing accelerations remove the degeneracy. Thus, a major source of systematic uncertainty would be removed allowing, in principle, for  accurate and genuine constraints on the physical parameters responsible of the post-Keplerian effects. Astrometric measurements of what could be identified with the sidereal period might be feasible in the near future with the ongoing GAIA mission for relatively detached exoplanets, perhaps characterized by $a > 0.5~\textrm{au}$. To this aim, it is interesting to remark how the general relativistic corrections are either independent of the size of the orbit  or grow as $\sqrt{a}$, while the quadrupolar ones fall as $1/\sqrt{a}$.
\section{Conclusions}\lb{conclusioni}
We analytically worked out the post-Keplerian corrections to some orbital periods which are degenerate in the purely Keplerian two-body scenario. We considered the Newtonian acceleration due to the oblateness $J_2$ of the primary, and the post-Newtonian general relativistic Schwarzschild-like and Lense-Thirring terms. We adopted a systematic perturbative approach which can be straightforwardly extended to any other disturbing acceleration, independently of its physical origin; as an example, it may find application in the orbital evolution of mass-transferring eccentric binary systems.  It turned out that the orbital periods considered differ from each other in presence of the aforementioned disturbing accelerations. We did not restrict to any particular orientation of the primary's spin axis  in  space, and no a-priori assumptions about the orientation  of the orbital plane with respect to the plane of the sky or the primary's equator were made.
The removal of such limitations, common to almost all the existing papers in the literature, is
important because the spatial orientation of the spin axes of astronomical
bodies of potential interest is often either imperfectly known or unknown at
all.  Also when the situation is more favorable, as in our Solar System, the direction of the planetary spin axes is known with a
necessarily limited accuracy. Finally, it must be remarked that, when data
analyses spanning several decades are involved, the fact that the spins of
many non-isolated bodies undergo slow precessional motions must be taken
into account as well.

In principle, the results obtained represent new tools to either constrain some key physical parameters of the primary and/or design new tests of general relativity in several astronomical and astrophysical systems, especially when no other means are available. It was demonstrated in the literature that it is possible to measure the draconitic period of an Earth's artificial satellite. Moreover, for the vast majority of the exoplanets so far discovered, for which no orbital precessions are accessible to observation, the strategy proposed here is able, in principle, to yield constraints on the parent star's oblateness which, otherwise, would not be possible to infer. To this aim, our theoretical results can  be used to identify those extrasolar systems for which post-Keplerian fits should be performed to extract unbiased orbital and physical parameters if the size of the predicted post-Keplerian  corrections is larger than the experimental accuracy in measuring the planetary
orbital periods.
If  it were possible to independently measure different periods for a given system, it would be possible to take their difference by canceling out the common Keplerian component, which represents a major source of systematic uncertainty, leaving just a post-Keplerian correction.

Our results could be applied also to binaries hosting astrophysical compact objects and to the stars orbiting the supermassive black hole in Sgr A$^\ast$. The continuous monitoring of all of such systems over the years with future facilities of increased accuracy will provide more and better constraints.
\section*{Acknowledgements}
I am grateful to J. N. Winn and M. Efroimsky for their careful reading of the manuscript and their important comments and remarks which greatly improved it.
I thank also K. Masuda for useful discussions.
\bibliography{Gclockbib}{}

\end{document}